\documentclass[twocolumn,pre,superscriptaddress]{revtex4}
\usepackage{tikz}
\usepackage{graphicx}%
\usepackage{color} 
\usepackage{transparent}
\usepackage{dcolumn}
\usepackage{amsmath}
\usepackage{amssymb, bm, wasysym}
\usepackage{latexsym}
\usepackage{hyperref} 
\usepackage{booktabs}
\usepackage{latexsym, bbold}
\begin{document}

\newcommand{\tikzcircle}[2][red,fill=red]{\tikz[baseline=-0.5ex]\draw[#1,radius=#2] (0,0) circle ;}%
\def\bea{\begin{eqnarray}}
\def\eea{\end{eqnarray}}
\def\beq{\begin{equation}}
\def\eeq{\end{equation}}
\def\f{\frac}
\def\k{\kappa}
\def\e{\epsilon}
\def\ve{\varepsilon}
\def\be{\beta}
\def\D{\Delta}
\def\h{\theta}
\def\t{\tau}
\def\a{\alpha}

\def\cDa{{\cal D}[X]}
\def\cD{{\cal D}[x]}
\def\cL{{\cal L}}
\def\cLo{{\cal L}_0}
\def\cLa{{\cal L}_1}

\def\Re{{\rm Re}}
\def\sj{\sum_{j=1}^2}
\def\rk{\rho^{ (k) }}
\def\rek{\rho^{ (1) }}
\def\cek{C^{ (1) }}
\def\rz{\rho^{ (0) }}
\def\rt{\rho^{ (2) }}
\def\rtb{\bar \rho^{ (2) }}
\def\trk{\tilde\rho^{ (k) }}
\def\trek{\tilde\rho^{ (1) }}
\def\trz{\tilde\rho^{ (0) }}
\def\trt{\tilde\rho^{ (2) }}
\def\r{\rho}
\def\tD{\tilde {D}}
\def\off{\w_{\rm off}}

\def\s{\sigma}
\def\kb{k_B}
\def\bF{\bar{\cal F}}
\def\F{{\cal F}}
\def\la{\langle}
\def\ra{\rangle}
\def\nn{\nonumber}
\def\up{\uparrow}
\def\dn{\downarrow}
\def\S{\Sigma}
\def\dg{\dagger}
\def\d{\delta}
\def\p{\partial}
\def\l{\lambda}
\def\L{\Lambda}
\def\G{\Gamma}
\def\o{\Omega}
\def\w{\omega}
\def\g{\gamma}

\def\jv{ {\bf j}}
\def\jr{ {\bf j}_r}
\def\jd{ {\bf j}_d}
\def\jdd{ { j}_d}
\def\noi{\noindent}
\def\a{\alpha}
\def\d{\delta}
\def\p{\partial} 

\def\la{\langle}
\def\ra{\rangle}
\def\e{\epsilon}
\def\n{\eta}
\def\g{\gamma}
\def\break#1{\pagebreak \vspace*{#1}}
\def\hf{\frac{1}{2}}

\title{Filament-motor protein system under loading:  instability and limit cycle oscillations}
\author{Amir Shee}
\email{amir@iopb.res.in}
\affiliation{Institute of Physics, Sachivalaya Marg, Bhubaneswar 751005, India}
\affiliation{Homi Bhaba National Institute, Anushaktigar, Mumbai 400094, India}

\author{Subhadip Ghosh}
\email{sghosh@phy.hr}
\affiliation{Department of Physics, Faculty of Science, University of Zagreb, Bijeni{\^c}ka cesta 32, 10000 Zagreb, Croatia}

\author{Debasish Chaudhuri}
\email{debc@iopb.res.in}
\affiliation{Institute of Physics, Sachivalaya Marg, Bhubaneswar 751005, India}
\affiliation{Homi Bhaba National Institute, Anushaktigar, Mumbai 400094, India}

\date{\today}

\begin{abstract}
We consider the dynamics of a rigid filament in a motor protein assay under external loading. The motor proteins are modeled as active harmonic linkers with tail ends immobilized on a substrate. Their heads attach to the filament stochastically to extend along it, resulting in a force on the filament, before detaching. The rate of extension and detachment are load dependent. Here we formulate and characterize the governing dynamics in the mean field approximation using linear stability analysis, and direct numerical simulations of the motor proteins and filament. Under constant loading, the system shows transition from a stable configuration to instability towards detachment of the filament from motor proteins. Under elastic loading, we find emergence of stable limit cycle oscillations via a supercritical Hopf bifurcation with change in activity and the number of motor proteins.  Numerical simulations of the system for large number of motor proteins show good agreement with the mean field predictions. 
\end{abstract}

\maketitle
\section{Introduction}
Cytoskeletal filaments and associated motor proteins (MP) stabilize structure of the cell and determine its dynamics~\cite{Alberts2009, Howard2001}. The cross-linking MPs, while extending towards one end of the polar filament, can shear filament pairs against each other, in an active non-equilibrium process hydrolyzing ATP. Within a living cell, the filaments form a meshwork, in which each filament encounters forces due to its surrounding~\cite{Jansen2015, Koenderink2009, Gardel2004}. While a major contribution to this force comes from active processes~\cite{Braun2016}, recent studies showed that entropic effects like that of depletion, and diffusible passive cross-linkers can lead to significant sliding forces on overlapping filaments~\cite{Schnauss2016, Walcott2010, Lansky2015, Ghosh2017}.

The gliding motion of filaments on a motor protein assay has been used extensively to study dynamics of cytoskeletal filaments outside the living cell. The competition between opposing groups of MPs can lead to spontaneous oscillations in gliding assays~\cite{Leduc2010}. 
Filament motion under cooperative MPs and position dependent load that could arise from passive cross-linkers or harmonic trap showed emergence of stable limit cycle oscillations~\cite{Ghosh2017, Placais2009, Julicher1997,Julicher1997a}.  
Similar spontaneous oscillations have been observed in many contexts in cell biology~\cite{Kruse2005a, Beta2017}, e.g., sarcomere oscillations, mitotic spindle oscillations, and chromosome oscillations~\cite{Gunther2007, Grill2005, Campas2006, Chikashige1994}.

Using mean field theory and stochastic simulations, we consider the motion of a filament in a gliding assay of MPs, in the presence of an external force.  As has been shown recently, the depletion potential in filament bundles can  change from a linear to harmonic form with increase in filament number~\cite{Schnauss2016}. We consider an external force that could be constant or be a function of filament position. Under a constant load, the filament on MP assay shows a dynamical crossover from stable to unstable phase. Whereas, in the presence of an elastic loading, the filament shows stable limit cycle oscillations when the number of MPs  is larger than a critical value~\cite{Grill2005, Placais2009}. We show how the onset of spontaneous oscillations depends on the MP activity in terms of its extension rate and detachment force, which can be tuned, e.g., by changing ATP concentration~\cite{Spudich1983, Schnitzer2000, Mizuno2007}. 

We present a linear stability analysis of mean field equations to find phase diagrams showing linearly stable and unstable phases, separated from stable and unstable spirals. This is compared them with numerical solutions of the non-linear equations. The boundary between the unstable spiral and linear instability disappears once nonlinearities are considered, and the whole region shows stable limit cycle oscillations. We show how the critical number of MPs required for the onset of spontaneous oscillations depend on the stiffness of the elastic load acting on the filament, a property that might be utilized by cells to sense stiffness of extra-cellular matrix. The mean field phase diagrams constitute our first main result. 

We present a derivation of the mean field equation for bound MPs using a Fokker-Planck approach, to identify the limitations in the approximation. However, a full numerical simulation of the model, distinguishing the individual MPs and incorporating the stochastic nature of the dynamics,  shows good agreement with the mean field prediction of the supercritical Hopf bifurcation boundary identifying the onset of spontaneous oscillations. This is our second main result. Deep inside the oscillatory phase, the dynamics shows a behavior typical of relaxation oscillators. The predictions of stable oscillations obtained from the mean field equations show good agreement with the stochastic simulations. In our numerical calculations, we use parameter values corresponding to microtubule and kinesin motor proteins, allowing our predictions amenable to direct experimental verification.

In Sec.~\ref{sec_model} we present our model. The linear stability analysis and its comparison with solutions of non-linear mean field equations are presented in Sec.~\ref{sec_mft}. In Sec.~\ref{sec_limcycle} we present the derivation of mean number of attached MPs using the Fokker-Planck approach. We also present the evolution of probability density of attached and detached fraction of MPs in this section. Results of numerical simulations of the detailed stochastic model and their comparison against the Fokker-Planck mean field approach is presented in Sec.~\ref{sec_simu}. Finally we conclude summarizing the main results in Sec.~\ref{sec_conc}.

\section{Model}
\label{sec_model}
\begin{figure}[t]
\begin{center}
\includegraphics[width=8cm]{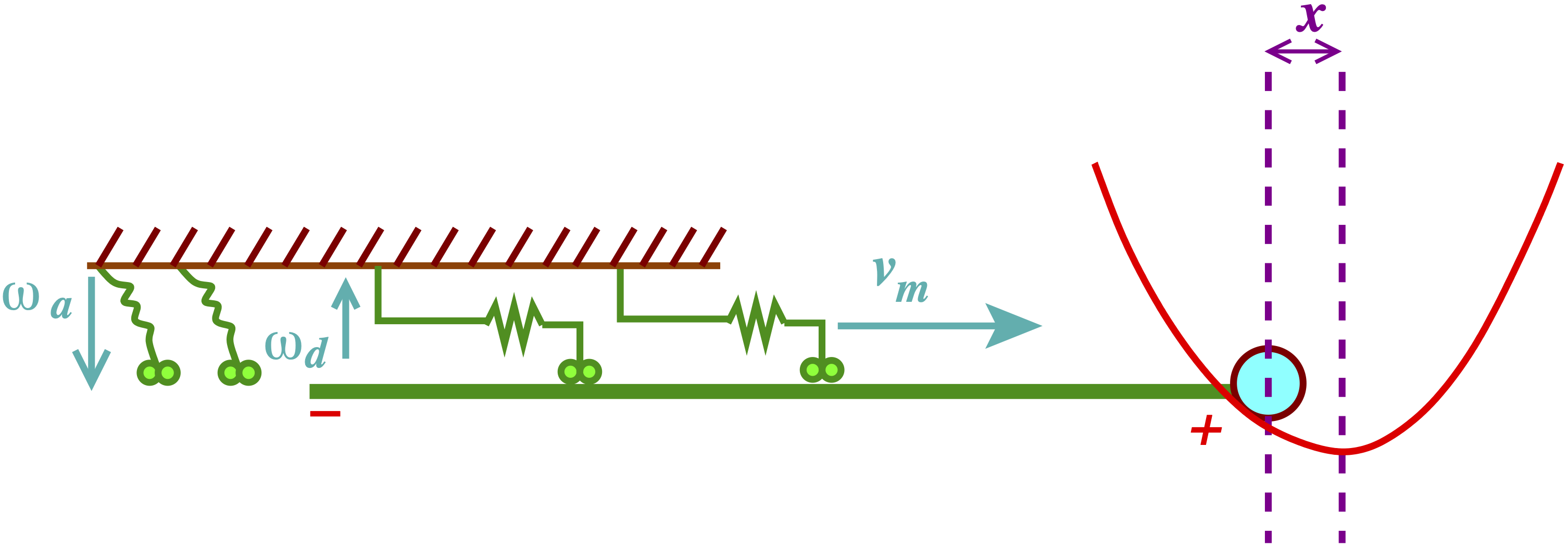}
\caption{(color online) 
Schematic diagram of the model where a motile MT filament is attached with a harmonic trap of stiffness constant $K_T$. When attached, kinesin walks along the filament towards right with a velocity $v_m$, pulling the filament towards left.}
\label{fig_MT_ext_harmonic}
\end{center}
\end{figure}

We consider a gliding assay set up~(Fig.\ref{fig_MT_ext_harmonic}) in which the {\em tail end} of the MPs are attached irreversibly to a cover slip. The MPs are assumed to be active harmonic linkers having stiffness $k_m$. The {\em head end} of MPs can attach to a segment of rigid filament floating on the assay within a cutoff range $r_c$ with a rate $\w_a$ in a diffusion limited manner.  The maximum number of MPs that can attach to the  filament of length $L$ is $N=L\, \phi_{MP}$, where $\phi_{MP}$ is the linear density of MPs attached to the substrate. The attached head of each MP extends along the filament in a directed fashion, from negative to positive end of the filament. This active extension requires energy consumption from ATP hydrolysis that brings the system out of equilibrium.  The rate of extension in $i$-th MP is denoted by an active velocity $v_m^i$ that depends on the load force $f_l^i = k_m y^i$ exerted on the MP due to the extension $y^i$ itself. We consider a piece-wise linear relation~\cite{Carter2005,Leduc2010}
\bea
v_m^i(f_l^i) =
\left\{
\begin{array}{ll}
v_0 & \mbox{for}\;\;\; f_l^i \leq 0 \\
v_0  \left( 1-\frac{f_l^i}{f_s} \right)& \mbox{for}\;\;\; 0<f_l^i \leq f_b,\, f_b>f_s  \\
-v_{back}& \mbox{for}\;\;\; f_l^i >f_b \\
\end{array}
\right.
\label{eq_vm}
\eea
where $f_s$ denotes the stall force and $v_0$ stands for the intrinsic MP velocity. For a load force beyond stall,  $f_l \geq f_b > f_s$,  the velocity saturates to an extremely small negative value $v_{back}$~\cite{Leduc2010,Carter2005}, while  supportive loads do not affect the intrinsic MP motion. Assuming the MPs to be forming slip bonds, the load dependent detachment rate is expressed as $\off= \w_d \exp(|f_l^i|/f_d)$. The attachment detachment ratio breaks detailed balance. 

All the parameters $v_0$, $\w_a$, $\w_d$, $f_s$, and  $f_d$ characterizing MPs are potentially functions of the ATP concentration in the ambient fluid. An assumption of Michaelis-Menten kinetics of ATP hydrolysis has been used to describe the  ATP dependence of $v_0$ for  kinesin, where $v_0$ increases linearly for small ATP concentrations to eventually saturate~\cite{Schnitzer2000, Chaudhuri2016}. Previous analysis of kinesin run-lengths demonstrated the ATP dependence of $f_d$~\cite{Schnitzer2000, Ghosh2017}. A change in $v_0$ leads to various interesting dynamical regimes. We return to this point later in the paper.

The over-damped dynamics of the filament position $x$ is determined by the mechanical force balance, 
\begin{equation}
 \g_f\dot{x}= F_m + F_e 
 \label{eq_xdot}
\end{equation}
where the left hand side corresponds to the  friction force characterized by $\g_f$ and associated to the relative motion of the filament $\dot x := dx/dt$ with respect to the substrate. The $n_m$ number of attached motor proteins exert a total force $F_m = -\sum_{i =1}^{n_m}  f_l^i$, and $F_e$ denotes the external loading that acts against the drive of the MPs.  
The filament motion can in turn drag the attached MPs along with it, such that the extension of $i$-th MP is given by
\bea
\dot{y}^i = v_m^i(f_l^i) + \dot{x}.
\label{eq_ydot}
\eea

\section{Results}
In this section we first present a mean field description of the model presented above. We utilize it to obtain linear stability predictions for dynamical phases and phase transitions in the presence of external loading on the filament. Numerical solutions of the non-linear mean field equations are used to compare with the linear stability results. In the second part, we present a Fokker-Planck description and derive the mean field equations to discuss its limitations. Finally, we perform detailed numerical simulations of the full stochastic model, and compare the results with mean field predictions.     
 
\subsection{Mean field theory}
\label{sec_mft}
We assume all the MPs to be equivalent within the mean field approximation, and describe them using the same average extension $y = (1/{n_m})\sum_{i=1}^{n_m} y^i$, where $n_m$ denotes the number of attached MPs.
To express the equations in a dimensionless form we use the energy scale set by $\kb T$, the time scale $\omega_{d}^{-1}$, and the length scale $l_0 = \sqrt{\kb T/\g_f \w_d}$. The unit of force is set by $f = \sqrt{k_B T \g_f \omega_d}$. Within mean field approximation, the dynamics is described by three coupled non-linear differential equations for dimensionless forms of filament position $\tilde x = x/l_0$, mean extension of MPs $\tilde y = y/l_0$, and the attached fraction of MPs $\tilde n_m=n_m/N$, 
\bea
\frac{d \tilde{x}}{d \tau} &=& \tilde{F}_e-N\tilde{n}_m \tilde{k}_m \tilde{y}, \nn\\
\frac{d \tilde{y}}{d \tau} &=& \tilde{v}_0\left(1-\frac{\tilde{k}_m \tilde{y}}{\tilde{f}_s}\right) + \frac{d \tilde{x}}{d \tau}, \nn\\
\frac{d \tilde{n}_m}{d \tau} &=& (1-\tilde{n}_m)\tilde{\omega} -  \tilde{n}_m \exp\left[\frac{\tilde{k}_m \tilde{y}}{\tilde{f}_d}\right]. 
\label{eq_mfLange}
\eea
In the above equations we used the dimensionless time $\tau = t \omega_d$, spring constant of MPs 
$\tilde{k}_m = {k_m l_0}/{f}$, attachment ratio $\tilde{\omega} = {\omega_a}/{{\omega_d}}$, stall force $\tilde{f}_s = {f_s}/{f}$, and detachment force $\tilde{f}_d = {f_d}/{f}$. In the presence of an external  load acting against directed MPs, the mean MP extension remains positive. This allows us to express the mean detachment rate as $\off = \w_d\exp(k_m y/f_d)$. We return to this point in Sec.~\ref{sec_limcycle}.   

In numerical estimates throughout this paper, we use parameter values typical of microtubule-kinesin assays shown in Table-\ref{table1}.  These values set the unit of length $l_0 = \sqrt{\kb T/\g_f \w_d}= 33$\,nm, force $f=\kb T/l_0=0.125\,$pN, and velocity $v=l_0\w_d = 33\,$nm/s. 
In the following, we first perform a linear stability analysis of Eq.(\ref{eq_mfLange}) using a constant loading $\tilde F_e$. 

\begin{table}[h]
\caption{Parameters: Two values of $v_0$ and $f_d$ correspond to ATP concentrations of $5\, \mu$M and 
2\,mM respectively.}
\centering
\begin{tabular}{l l l}
\hline\hline
active velocity & $v_0$ & 0.006, 0.8$\mu$m/s~\cite{Schnitzer2000,Block2003}\\
stall force & $f_s$ & 7.5 pN~\cite{Carter2005,Block2003} \\
back velocity & $v_{back}$ & 0.02$\mu$m/s~\cite{Carter2005} \\
detachment force & $f_d$ & 1.8, 2.4 pN~\cite{Ghosh2017}\\
attachment rate & $\w_a$ &  5, 20/s~\cite{Scharrel2014, Leduc2004, Schnitzer2000} \\
detachment rate & $\w_d$ & 1/s~\cite{Block2003} \\
motor stiffness & $k_{m}$ & 1.7 pN/nm~\cite{Ghosh2017}~\footnote{same order of magnitude as in Ref.~\cite{Dogterom2005,Grill2005}}, 0.3 pN/nm~\cite{Kawaguchi2001}\\ 
MT viscous friction & $\g_f$ & 893 $\kb$T-s/$\mu$m$^2$~\cite{Lansky2015} \\
\hline\hline
\end{tabular}
\label{table1}
\end{table}%

\begin{figure}[!t]
\begin{center}
\includegraphics[width=8cm]{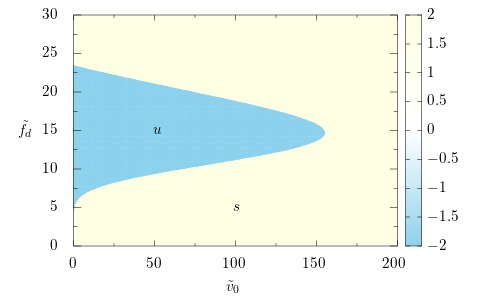}
\caption{ (color online) 
Phase diagram under constant loading in $\tilde{v}_0-\tilde{f_d}$ plane shown using the heat map of the quantity $q$ defined in Eq.(\ref{eq_q}). We use parameter values typical of a microtubule- kinesin system, $\tilde k_m = 450$\,($k_m=1.7\,$pN/nm), $\tilde{f}_s = 60$, $\tilde{\omega} = 20$ and use $N = 5$ number of MPs. The color-box shows the mapping of $q$-values to the color code. The light blue (yellow) region denotes unstable (stable) phase under perturbation. }
\label{fig_cons_force_pd}
\end{center}
\end{figure}

\subsubsection{Constant loading}
 The fixed points of the system of equations is obtained by setting all the time derivatives of Eq.(\ref{eq_mfLange}) to zero to obtain,
\bea
\tilde{k}_m \tilde{y}_0 = \tilde{f}_s,~ 
\tilde{n}_m^0 = {\tilde{\omega}} / [{\tilde{\omega}+\exp({\tilde{f}_s}/{\tilde{f}_d})} ] 
= \tilde F_e/N \tilde f_s,
\eea
The last relation determines the loading $\tilde F_e$ corresponding to the fixed point of a system of $N$ MPs characterized by attachment ratio $\tilde \w$, stall force $\tilde f_s$ and detachment force $\tilde f_d$. 
In the absence of position dependence of the external force, the fixed point is  $\tilde{x}$- independent.  
Its stability can be analyzed considering the evolution $d | \psi \ra/d\t  =$ {\boldmath $a$} $| \psi \ra$ of a small perturbation $|\psi \ra = (\d \tilde x, \d \tilde y, \d \tilde n_m)$,  where {\boldmath $a$} denotes the stability matrix with elements
\begin{align}
& a_{11} = 0, 
a_{12} = -\tilde{k}_m N \tilde{n}_m^0, 
a_{13} = -\tilde{f}_s N, 
a_{21} = 0, \nn\\ 
& a_{22}= -\Big( \tilde \mu + \tilde{k}_m N \tilde{n}_m^0\Big),
a_{23} = a_{13},\, a_{31} = 0, \nn\\
& a_{32} = -(\tilde{k}_m / \tilde{f}_d) \tilde \w (1-\tilde n_m^0), 
a_{33} = - \tilde \w/ \tilde n_m^0, \nn
\end{align}
where $\tilde \mu := \tilde v_0 /\tilde y_0$. 
Diagonalizing the stability matrix {\boldmath $a$} gives the characteristic equation 
$\lambda(\lambda^2 + p\lambda + q) = 0$ 
with solutions $\l_1=0$, and the other two eigenvalues given by $\lambda_{\pm} = (1/2)[-p\pm\sqrt{p^2 - 4q}]$. Here $p=-(a_{22} + a_{33})>0$ always, and 
\bea
q=(a_{22}a_{33} - a_{13}a_{32})
\label{eq_q}
\eea
may change sign, thereby controlling the stability of the fixed point. In this case, the smaller eigenvalue $\l_-$ remains negative, however, $\l_+$ can become positive when $q$ changes sign from positive to negative. Thus $q=0$ line denotes the boundary between stable and unstable fixed points. This phase boundary is shown in Fig.~\ref{fig_cons_force_pd}.
{
In the stable phase, the force balance is maintained as the filament moves in a direction opposite to the extension of MPs. In the unstable phase, the extension of MPs cannot stabilize the filament position, which slides in the direction of extension of the attached MPs.}  
However, since the discriminant $p^2-4q = (a_{22} - a_{33})^2 + 4a_{13}a_{32} > 0$, the quadratic equation does not support any imaginary part in the eigenvalues. Thus oscillatory behavior, stable or unstable, is ruled out. Under constant external loading the filament driven by MPs cannot sustain oscillations.


\subsubsection{Elastic loading}
For elastic loading $\tilde F_e = - \tilde{K}_T\tilde{x}$. This might be generated on the filament by trapping one of its ends by a laser tweezer or atomic force microscope tip. 
As in the previous case, we perform a linear stability analysis. The fixed points of the mean field dynamics is given by
\begin{eqnarray}
\tilde{k}_m \tilde{y}_0 = \tilde{f}_s,~  
\tilde{n}_m^0 = {\tilde{\omega}} / [{\tilde{\omega}+\exp({\tilde{f}_s}/{\tilde{f}_d})} ],~ 
\tilde{x}_0 = -\frac{\tilde{n}_m^0N\tilde{f}_s}{\tilde{K}_T}. \nn
\end{eqnarray}
The small perturbations around the fixed point $|\psi \ra = (\d \tilde x, \d \tilde y, \d \tilde n_m)$ evolves with $d | \psi \ra/d\t  =$ {\boldmath $a'$} $| \psi \ra$ where the elements of the stability matrix {\boldmath $a'$} are given by
\begin{align}
& a_{11}^{\prime} = -\tilde{K}_T, 
a_{12}^{\prime} = -\tilde{k}_m  \tilde{n}_m^0\, N, \nn\\
& a_{13}^{\prime} = -\tilde{f}_s N, 
a_{21}^{\prime} = a_{11}^{\prime}, \nn\\
& a_{22}^{\prime}= -(\tilde \mu + \tilde{k}_m N \tilde{n}_m^0),
a_{23}^{\prime} = a_{13}^{\prime},\, a_{31}^{\prime} = 0, \nn\\
& a_{32}^{\prime} =  -(\tilde{k}_m / \tilde{f}_d) \tilde \w (1-\tilde n_m^0),  
a_{33}^{\prime} = - \tilde \w / \tilde n_m^0,
\end{align}
where, as before, $\tilde \mu := \tilde v_0/ \tilde y_0$. 
In the case of constant loading, the matrix element $a_{11}$ was zero, reducing one eigenvalue to zero. The other two eigenvalues were determined by a quadratic equation. However, for elastic loading, $a'_{11} \neq 0$ and  
the eigenvalues are given by the full cubic equation 
\begin{equation}
\lambda^3 + A\lambda^2 + B\lambda + C = 0
\label{eq:eigen_harmonic}
\end{equation}

In terms of different matrix elements $A = -$Tr({\boldmath $a'$}), $B = \hf\left(a_{ii} a_{jj} -a_{ij} a_{ji} \right)$ where we implied summation over repeated indices, and $C=-$det({\boldmath $a'$}).
They can be expressed as 
\bea
A &=&  \tilde \mu + \tilde{K}_T  + \tilde \w / \tilde n_m^0 +  \tilde{n}_m^0 \tilde{k}_m\, N \nn\\
 B &=& \tilde \mu \tilde{K}_T  +  \f{\tilde \w}{ \tilde n_m^0} (\tilde \mu + \tilde{K}_T) 
 +\tilde k_m \tilde \w \left[ 1 - \f{\tilde f_s}{\tilde f_d} (1-\tilde n_m^0 ) \right] N\nn\\
C &=& \tilde \mu \tilde{K}_T \, \tilde \w / \tilde n_m^0.
\label{eq:harmonic_coeff}
\eea 
Properties of these coefficients determine the existence of different phases and the dynamical behavior of the system. A cubic polynomial has eight possible combinations of real and complex roots. Eq.(\ref{eq:harmonic_coeff}) shows that $A$ is positive definite, $C$ is positive semi definite, while $B$ can change its sign. These strong restrictions eliminate four combinations for roots to the cubic polynomial $\lambda_{1,2,3}$. The remaining four combinations characterize the four different phases in the system. The possible combinations are as follows:
$(i)$~All three eigenvalues $\lambda_{1,2,3}$  are real negative, characterizing a {\em linearly stable} $(s)$ phase. 
$(ii)$~$\lambda_1$ is real negative, but $\lambda_{2,3}$ are real positive characterizing a {\em linearly unstable} $(u)$ phase. 
$(iii)$~$\lambda_1$ is real negative. On the other hand $\lambda_{2,3}$ are complex conjugate pairs with negative real part, $\lambda_{2,3} =-\alpha\pm i\beta$, characterizing a decaying oscillation of perturbations in {\em stable spiral} $(ss)$ phase. 
$(iv)$~$\lambda_1$ is real negative though $\lambda_{2,3}$ are complex conjugates with positive real part. $\lambda_{2,3} =\alpha\pm i\beta$ with $\a>0$ denotes oscillations with growing amplitude in {\em unstable spiral} $(us)$ phase. 

\begin{figure}[!t]
\begin{center}
\includegraphics[width=8cm]{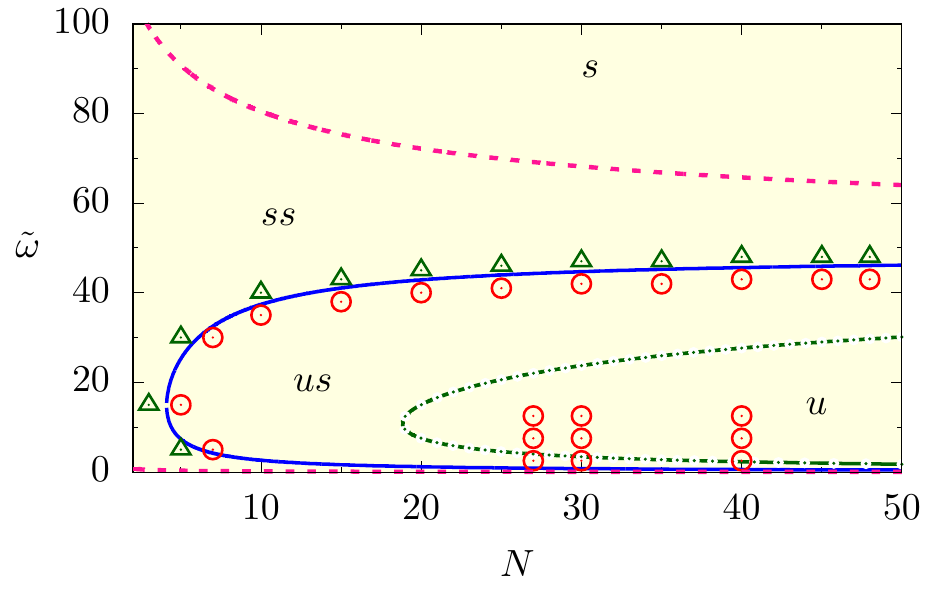}  
\caption{ 
Phase diagram for filament in MP assay under elastic loading in $N-\tilde{\omega}$ plane, with $\tilde k_m = 450$, $\tilde{v}_0 = 24.24$, $\tilde{f}_s = 60$, $\tilde{f}_d = 19.2$, $\tilde K_T = 83$ kept fixed. 
The lines denote the linear stability phase boundaries between linearly stable ($s$), stable spiral ($ss$), unstable spiral ($us$), and linearly unstable ($u$) phases.   The points denoted by $\triangle$ and $\Circle$ indicate decaying oscillations and limit cycle oscillations, respectively, corresponding to the full non-linear dynamics in Eq.(\ref{eq_mfLange}). 
}
\label{fig_harmonic_N_Wco}
\end{center}
\end{figure}

{\bf Phase transitions:} ($a$)~{\em Phase boundary between linearly (un)\,stable and (un)\,stable spiral phase}: The complex conjugate roots disappear as the minimum of the polynomial $p(\l) = \l^3+A \l^2+ B\l$ touches the line $p(\l)=-C$, corresponding to two degenerate eigenvalues. 
The minimum of $p(\l)$ is at $\lambda_m = -\frac{A}{3} + \frac{1}{3}\sqrt{A^2 - 3B}$. This condition lead to the phase boundary,
\bea
C = \left[\frac{A}{3} + \frac{2}{3}\sqrt{A^2 - 3B}\right]\left[-\frac{A}{3} + \frac{1}{3}\sqrt{A^2 - 3B}\right]^2 
\label{eq_pb_sta_unsta}
\eea 
For $B \geq 0$, the boundary is between linear stable ($s$) and stable spiral ($ss$) phases. On the the hand, for $B\textless0$, it denotes the boundary between unstable spiral ($us$) and linearly unstable ($u$) phases.

{\em ($b$)~Phase boundary between stable spiral and unstable spiral phases}: As the sign of the real part $\a$ of complex conjugate roots $\l_{2,3}=\a\pm i\be$ changes from negative to positive the system becomes unstable and start to oscillate. This transition is captured by setting $\a=0$. This leads to the condition 
\begin{equation}
C - AB = 0
\label{eq_pb_os_sta_unsta}
\end{equation}
The growing amplitudes of oscillations in $us$ phase predicted by linear stability analysis gets stabilized  by non-linearities into stable limit cycle oscillations, as we show later in Sec.~\ref{sec_limcycle} using full stochastic simulations and further analysis. This denotes a supercritical Hopf bifurcation to a stable limit cycle, e.g., at a critical number of MPs $N^\ast$~(Fig.\ref{fig_harmonic_N_Wco}). The expression of $N^\ast$ can be obtained by solving Eq.(\ref{eq_pb_os_sta_unsta}), a quadratic equation in terms of $N^\ast$. Note that in the absence of elastic loading $K_T=0$ the polynomial coefficient $C=0$. As a result Eq.(\ref{eq_pb_os_sta_unsta}) cannot be satisfied, and the Hopf bifurcation disappears.   

At the supercritical Hopf bifurcation, the imaginary part of the eigenvalue $\be = \sqrt{B}$, so that the MP- filament system shows oscillations with a frequency $f_\w = \sqrt{B}/2\pi$. Clearly, the frequency of oscillations  $f_\w$ depends on the number of MPs, ATP-dependent activity of MPs determined by active velocity, attachment detachment rates, stall force, and detachment force, apart from the effective elastic constant of the external loading force.

\begin{figure}[!t]
\begin{center}
\includegraphics[width=8cm]{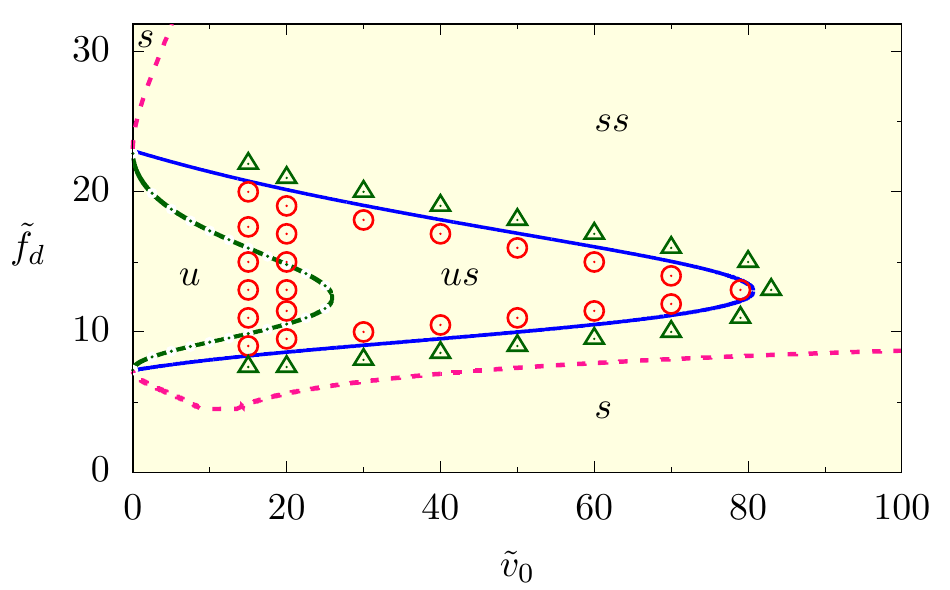}  
\caption{ 
Phase diagram for filament in MP assay under harmonic trap in $\tilde{v}_0-\tilde{f}_d$ plane, keeping $\tilde k_m = 450$, $\tilde{\omega} = 20$, $\tilde{f}_s = 60$, $N = 5$ and $\tilde{K}_T = 83$ fixed. The phases and phase boundaries are indicated in the same manner as in Fig.\ref{fig_harmonic_N_Wco}. 
}
\label{fig_harmonic_v0_fd}
\end{center}
\end{figure}

{\bf Phase diagrams:} Using Eq.(\ref{eq_pb_sta_unsta}) and Eq.(\ref{eq_pb_os_sta_unsta}) we present two phase diagrams showing the transitions between the above mentioned phases in Fig.~\ref{fig_harmonic_N_Wco}, and \ref{fig_harmonic_v0_fd}. We use parameter values corresponding to kinesin-microtubule assays~(Table-\ref{table1}). 

In Fig.~\ref{fig_harmonic_N_Wco}, we show the phase diagram in number of MPs $N$- and attachment ratio $\tilde{\omega}$ plane.  This  indicates the requirement of a threshold number of MPs $N^\ast$ to get sustained oscillations in the $us$ phase. In addition, sustained oscillations depend on the activity in the system, parametrized in terms of attachment-detachment ratio $\tilde \w$, active velocity $\tilde v_0$ and the detachment force $\tilde f_d$. The lines in the plot show phase boundaries obtained from linear stability analysis, using Eq.s~(\ref{eq_pb_sta_unsta}) and (\ref{eq_pb_os_sta_unsta}). The boundary between $ss$ and $us$ phase appears via a supercritical Hopf bifurcation. The full non-linear dynamics corresponding to Eq.(\ref{eq_mfLange}) show decaying oscillations in $ss$ phase ($\triangle$), and limit cycle oscillations ($\Circle$) in the region denoted by $us$ and $u$. Clearly, once the non-linearities are considered, the boundary between $us$ and $u$ phase become irrelevant, the whole region inside the $us$ boundary shows limit cycle oscillations.

In Fig.~\ref{fig_harmonic_v0_fd} we further characterize these dynamical phase transitions in terms the of detachment force $\tilde f_d$ and active velocity $\tilde v_0$.  
As before, the linear stability analysis shows boundaries between $s$, $ss$, $us$ and $u$ phases. The consideration of non-linearities show that the whole region of $us$ and $u$ display limit cycle oscillations. The stable limit cycle phase appears from stable spiral via a supercritical Hopf bifurcation. This transition will be explored in further detail in Sec.~\ref{sec_simu} using numerical simulations of the stochastic dynamics governing the MPs and filament. 
Note that the phase boundaries between $ss$ and $s$ in Fig.s~\ref{fig_harmonic_N_Wco} and \ref{fig_harmonic_v0_fd}  are inconsequential, as both the phases are stable in long time limit. 

\begin{figure}[!t]
\begin{center}
\includegraphics[width=8cm]{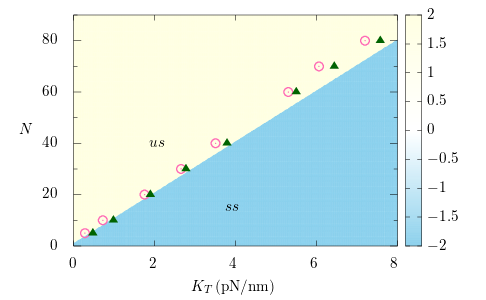}  
\caption{ 
The linear stability phase boundary between the stable spiral (blue: $ss$) and unstable spiral (yellow: $us$) phase in the plane of elastic loading stiffness $K_T$ and MP number $N$ is shown using the heat map of $C-AB$ in Eq.(\ref{eq_pb_os_sta_unsta}). The color box shows the mapping for the values of the function. Parameters used correspond to kinesin-microtubule assay, keeping $\tilde k_m = 450$, $\tilde{\omega} = 20$, $\tilde{f}_s = 60$, $\tilde f_d=19.2$, $\tilde v_0=24.24$ fixed. The points denoted by $\triangle$ and $\Circle$ indicate decaying oscillations ($ss$) and stable limit cycle oscillations ($us$), respectively, corresponding to the full non-linear dynamics shown in Eq.(\ref{eq_mfLange}).  Here we express $K_T$ in units of pN/nm. 
}
\label{fig_rigidity}
\end{center}
\end{figure}

In Fig.~\ref{fig_rigidity}, we show how the onset of stable limit cycle oscillations ($us$) depends on the number of MPs $N$ recruited for a given rigidity $K_T$ of the elastic loading. The plot uses parameter values corresponding to microtubule-kinesin MP assay, at an ATP concentration of $2$\,mM. The minimum number of MPs required for the onset of spontaneous oscillations increases with the stiffness $K_T$ of the substrate. While the particular calculations are performed for microtubule-kinesin system, the physical mechanism is equally applicable for acto-myosin systems. 
Our simple setup has a parallel in the rigidity sensing by cells, where contractile acto-myosin system couples to the extra-cellular matrix (ECM) via an adhesion complex consisting of alpha-actinin and integrin~\cite{Wolfenson2016, Lohner2019}. The range of $K_T$ values used in  Fig.~\ref{fig_rigidity} belongs to the range of rigidities of sub-micron elastomeric pillars used in cell spreading  experiments~\cite{Lohner2019}. 
The cell may utilize an increase of processive myosin bundles, required for the onset of oscillations (tugging), as a strategy to sense the ECM stiffness~\cite{Lohner2019, Plotnikov2012}. In fact, larger multifilament assemblies of myosin is noted near more rigid substrate~\cite{Lohner2019}.

The parameter values used in the above phase diagrams correspond to a gliding assay of microtubule on kinesin MPs~(Table-\ref{table1}).    
The elastic loading on the filament can be applied by optical tweezers or atomic force microscopes~\cite{Keya2017,  Placais2009, Kawaguchi2001}. 
In Fig.~\ref{fig_harmonic_N_Wco} and \ref{fig_harmonic_v0_fd}, we used a value of dimensionless stiffness $\tilde K_T$  that corresponds to $0.3$\,pN/nm. While the location of phase boundaries depends on $\tilde K_T$, the qualitative features remain unaltered. In the limit of extremely small $\tilde K_T$, however, the harmonic trap can act like a constant loading, as was shown in Ref.~\cite{Kreuzer2001}. Within the cell, our study has relevance for the relative sliding motion of filaments where the loading might be provided by other cellular components, or the filament bending~\cite{Grill2005}, and may have implications for rigidity sensing by cells~\cite{Wolfenson2016, Lohner2019}. We considered length stabilized filaments, which are typically used in gliding assay experiments, thus disregarding the possible effects of active polymerization depolymerization of filaments in  living cells~\cite{Howard2001}.

\subsection{Fokker-Planck approach to mean field}
\label{sec_limcycle}
Having established the phase diagrams using mean field equations and linear stability analysis, in this section we use a Fokker-Planck approach~\cite{Grill2005} involving the probability distributions $P_{a,d}(y,t)$ of attached and detached fractions of MPs to derive and analyze the mean field equations. The distribution functions obey the normalization $\int_{-\infty}^{\infty}(P_{a}+P_{d})\,dy = 1$, and evolve as  
\bea
\partial_{t} P_{a}+\partial_{y}J_{a} &=& \omega_{a} P_{d} -\w_d P_{a}\nn\\
\partial_{t} P_{d}+\partial_{y}J_{d} &=& -\omega_{a} P_{d} +\w_d P_{a},
\label{eq_fp} 
\eea
where, the probability currents
\bea
J_{a} &=& \dot{y} P_{a}(y,t)-D_{a}\,\partial_{y}P_{a}(y,t)\nn\\
J_{d} &=&-\nu y P_{d}(y,t)-D_{d}\,\partial_{y}P_{d}(y,t).
\label{eq_j}
\eea
Here $D_{a}$, $D_{d}$ are the diffusion coefficient of attached and detached MPs, respectively, and $\nu$ is the relaxation rate of the extension for the detached MPs. 

\begin{figure}[!t]
\begin{center}
\includegraphics[width=8cm]{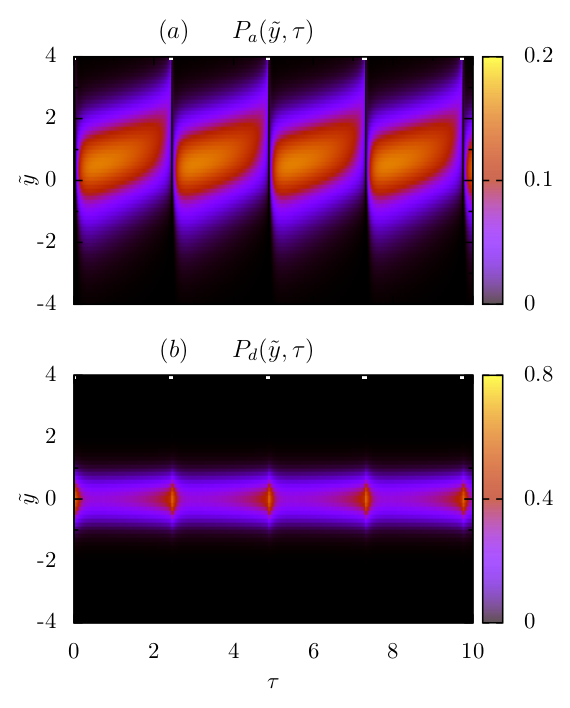}
\caption{ (color online) 
Kymographs show the time evolution of the probability distributions of MPs with extension $\tilde{y}$: ($a$)\,in the attached state\,($P_a$), and ($b$)\,in the detached state\,($P_d$). The color-box describes the values of probability distributions. The relaxation dynamics are determined by dimensionless diffusion constants $\tilde D_a=15.6$, $\tilde D_d = 17.4$, and the relaxation rate $\tilde \nu = 80$. Other  parameter values used are $\tilde{v}_0 = 50$, $\tilde{f}_d = 4$, $\tilde k_m = 4.57$, $\tilde{\omega} = 5$, $\tilde{f}_s = 14.29$, $N=160$ and $\tilde{K}_T = 4.57$.} 
\label{fig_Kymograph}
\end{center}
\end{figure}

As the typical relaxation rate is much faster than the attachment rate, $\nu \gg \w_a$, one can assume the detached MPs relaxes immediately to equilibrium distribution,
\bea
P_{d}(y,t)=\tilde n_d(t)\, A \exp\left(-\frac{k_m y^{2}}{2 k_{B}T}\right)
\label{eq_Pd}
\eea
with normalization $A=(k_m/2\pi \kb T)^{1/2}$. 
As a result, the fraction of detached MPs $\tilde n_d(t) = \int_{-\infty}^\infty dy P_{d}(y,t)$. This is related to the fraction of attached MPs $\tilde n_m(t) =  \int_{-\infty}^\infty dy P_{a}(y,t)$ via the conservation of total probability $ \tilde n_m(t) =1-\tilde n_d(t)$. 
Integrating the first equation of Eq.(\ref{eq_fp}) we get
\bea
\f{d \tilde n_m}{dt} = (1-\tilde n_m) \w_a - \la \w_d(y) \ra \tilde n_m
\eea
where $ \la \w_d(y) \ra := \int_{-\infty}^\infty dy P_{a}(y,t) \w_d(y)/\int_{-\infty}^\infty dy P_{a}(y,t)$. Using the expression for slip bond $\w_d(y) = \w_d \exp(k_m |y|/f_d)$ leads to 
\bea
\f{d \tilde n_m}{dt} = (1-\tilde n_m) \w_a - \w_d \la e^{k_m |y|/f_d} \ra \tilde n_m
\label{eq_nm_2}
\eea

By Jensen's inequality $\la e^{k_m |y|/f_d} \ra \geq e^{k_m \bar y/f_d}$ with $\bar y = \la |y| \ra$ denoting the mean extension in the attached state. Thus the actual relaxation 
${d \tilde n_m}/{dt}$ 
is slower than that  assumed in Eq.(\ref{eq_mfLange}).

The active extension and relaxation dynamics in terms of $P_{a,d}(y,t)$ is shown in Fig.\ref{fig_Kymograph} by numerically integrating the Fokker-Planck equations~(Eq.(\ref{eq_fp}) and (\ref{eq_j})\,) along with the evolution of filament position $x$~(Eq.(\ref{eq_xdot})\,), and mean extension of MPs following  
\bea
\dot y = v_m(y) + \dot x.
\label{eq_ydot_2}
\eea 
In the above equation the piecewise linear form of $v_m$ is used from Eq.(\ref{eq_vm}) replacing the load force $f_l = k_m y$. In Eq.(\ref{eq_xdot}), at this point, we use $F_m = -k_m \sum_{i=1}^{n_m} y^i = - k_m \int_{-\infty}^\infty dy\, y\, P_a(y,t) $. This reduces Eq.(\ref{eq_xdot}) to
\bea
\g_f \dot x = - K_T x - k_m \int_{-\infty}^\infty dy\, y\, P_a(y,t). 
\label{eq_xdot_2} 
\eea
In plotting this graph we used $D_d = \kb T/\g_f$, $D_a<D_d$, and $\nu=k_m/\g_f$. The last choice maintains $\nu \gg \w_a$. The parameters used correspond to kinesin-microtubule system~(Table-\ref{table1}). Here the unit of length is chosen to be $l_0=8\,$nm, the dimer-size of microtubules~\cite{Howard2001}. The unit of force is $f=\kb T/l_0 = 0.525\,$pN, and time is $\w_d^{-1}=1\,$s. As before, we express dimensionless extensions $\tilde y= y/l_0$, $\tilde x = x/l_0$ and dimensionless time $\t=\w_d t$. 

Due to the directed nature of the MP extension and the loading acting against MPs on an average, the mean extension $\tilde y$ of the attached fraction remains positive~(Fig.\ref{fig_Kymograph}($a$)\,), a fact used in replacing $\exp(k_m |\tilde y|/f_d)$ by $\exp(k_m \tilde y/f_d)$ in the mean field description of Eq.(\ref{eq_mfLange}). As Fig.\ref{fig_Kymograph}($a$) shows in terms of the evolution of $P_a(\tilde y, \t)$, the mean extension grows slowly up to a maximum, before relaxing back rapidly to zero in a time periodic manner. This is a characteristic of relaxation oscillators~\cite{Strogatz2014}. 
{
With detachment of one MP, the shared load on other attached MPs increases, increasing the effective rate of detachment. This mediates an avalanche of MP detachment leading to the rapid relaxation.  Once relaxed, MPs reattaches, maintaining oscillations.} 
The avalanche in detachments lead to an associated rapid increase in $P_d(\tilde y, \t)$~(Fig.\ref{fig_Kymograph}($b$)\,). Note that the distribution $P_d(\tilde y, \t)$ maintains a maximum at $\tilde y =0$, and is always symmetric around $\tilde y=0$, vindicating the simplification used in Eq.(\ref{eq_Pd}). 

\begin{figure}[!t]
\begin{center}
\includegraphics[width=8cm]{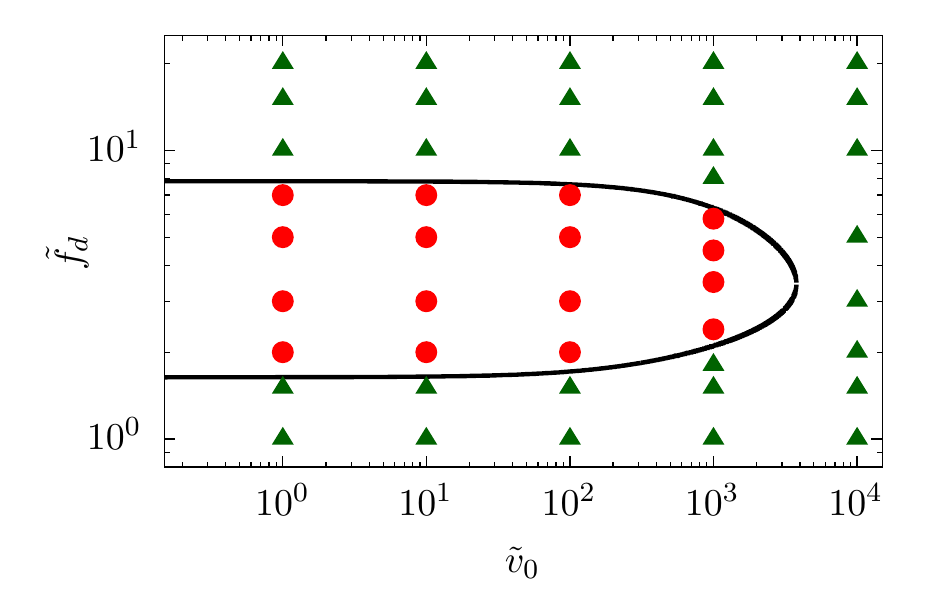}  
\caption{ (color online) 
Phase diagram for harmonically trapped microtubule-kinesin assay in $\tilde{v}_0-\tilde f_d$ plane, at fixed $\tilde k_m = 4.57$, $\tilde{\omega} = 5$, $\tilde{f}_s = 14.29$, $N=160$ and $\tilde{K}_T = 4.57$. The points denote the two phases characterized by decaying oscillations~($\triangle$) and stable limit cycle oscillations~($\Circle$) obtained from numerical simulations. The solid black line identifies the boundary of supercritical Hopf-bifurcation  predicted by Eq.(\ref{eq_pb_os_sta_unsta}). 
} 
\label{fig_Hopf_phdia}
\end{center}
\end{figure}

\subsection{Numerical simulations}
\label{sec_simu}
Finally, we perform a numerical simulation of the MP-filament model described in Sec.~\ref{sec_model}. We consider a kinesin-microtubule system with the microtubule held in its  positive end using a harmonic trap of strength $K_T$. A similar experimental system of myosin- F-actin with the F-actin held by an elastic load provided by a laser tweezer was considered before in Ref.~\cite{Placais2009}. We model the microtubule as a connected rigid string of $\s=8$\, nm segments. In this section, we use $l_0=\s$ as the unit of length, which sets the unit of force $f=\kb T/l_0 = 0.525\,$pN. We still use the unit of time $\w_d^{-1}=1\,$s. The $i$-th kinesin can attach to a microtubule segment within the cutoff radius $r_c/l_0=1.0$ stochastically with rate $\w_a$. When attached, the MP extends towards the plus end of  microtubule stochastically with a rate $v_m/l_0$. The instantaneous load force on the $i$-th MP is $f_l^i = k_m y^i$, expressed in terms of the extension $y^i(t)$. We use Eq.(\ref{eq_vm}) for the load dependence of the extension rate. The MPs detach from the filament stochastically with the rate $\w_d \exp(k_m |y^i|/f_d)$. This system along with Eq.s(\ref{eq_xdot}) and (\ref{eq_ydot}) for the position of the filament $x$ and MP extension $y_i$ are integrated numerically using Euler-Maruyama integration with time steps small enough so that probability of each event stays  less than one. We simulate a large number of MPs $N=160$. 

Performing the numerical simulations  over a range of activity $\tilde v_0$ and detachment force $\tilde f_d$ we obtain the phase diagram in Fig.~\ref{fig_Hopf_phdia}($a$). It shows two phases, one is characterized by decaying oscillations corresponding to stability~($\triangle$), and the other displays stable limit cycle oscillations~($\Circle$). The phase boundary obtained from numerical simulations of the full stochastic model shows good agreement with analytic result, Eq.(\ref{eq_pb_os_sta_unsta}).

\begin{figure}[!t]
\begin{center}
\includegraphics[width=8cm]{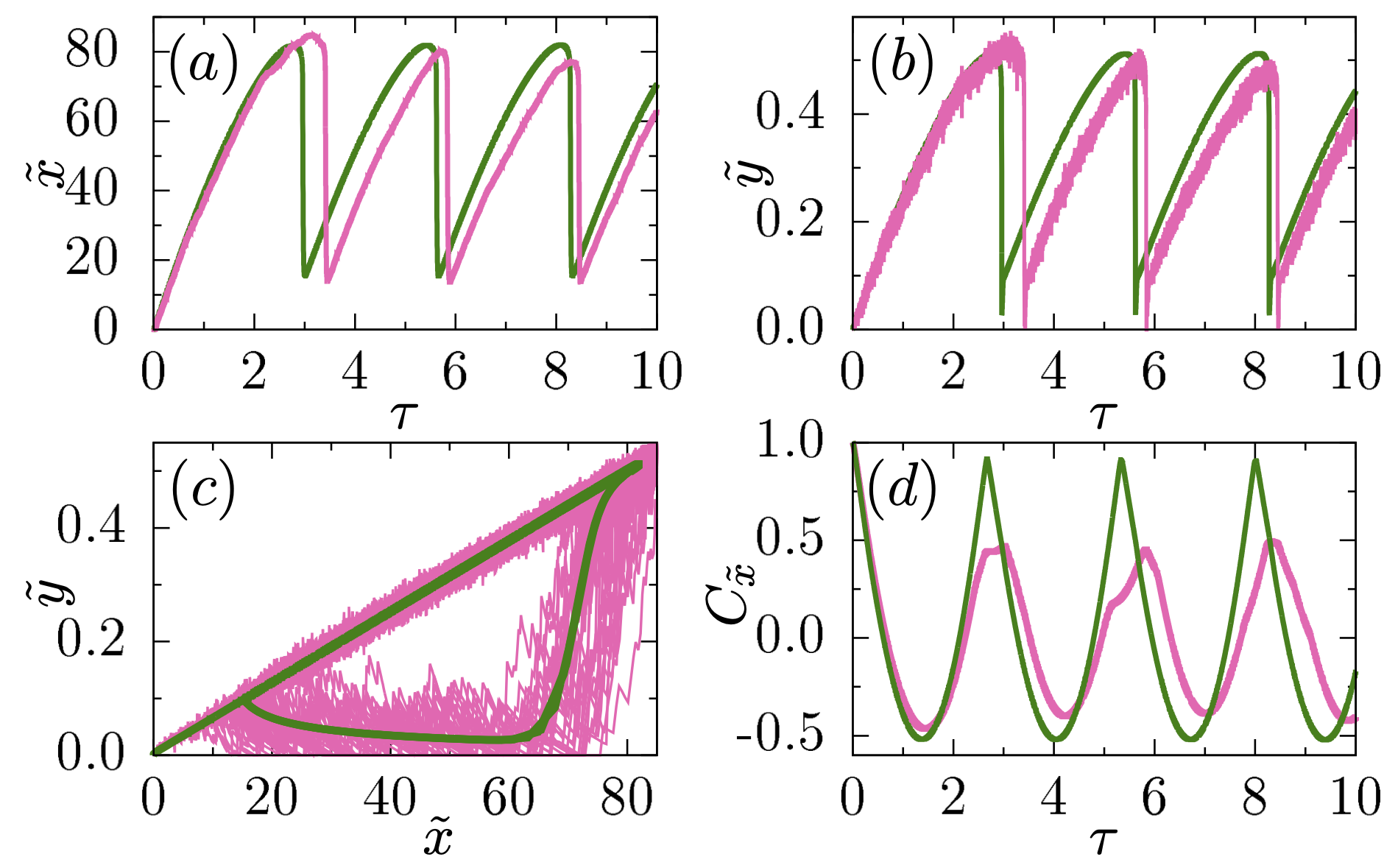}
\caption{ (color online) Time evolutions of $(a)$\,microtubule displacement $\tilde x(\t)$, and $(b)$\,mean kinsein extension $\tilde y(\t)$. ($c$)\,A parametric plot of $\tilde x(\t)$ and $\tilde y(\t)$ shows stable limit cycle. ($d$)\,Correlation function of microtubule displacement $C_{\tilde x}(\t)$ in the time-periodic steady state. The red (green) lines in all these figures correspond to numerical simulations (solutions to the Fokker-Planck based mean field equations). We used $\tilde{v}_0 = 50$, $\tilde{f}_d = 4$. All other parameter values are the same as in Fig.~\ref{fig_Hopf_phdia}. 
} 
\label{fig_evo}
\end{center}
\end{figure}

The dynamics deep inside the stable limit cycle phase is illustrated in Fig.~\ref{fig_evo}, using results from numerical simulations. We plot the evolutions of filament position $\tilde x(\t)$ and mean extension of kinesins $\tilde y(\t)$ deep inside the limit cycle phase ($\tilde v_0=50$, $\tilde f_d=4$). 
They show anharmonic oscillations with a well defined periodicity~(Fig.~\ref{fig_evo}($a$), ($b$)\,).
The slow extension (increase in $\tilde x$) followed by rapid relaxation is a typical relaxation oscillatior behavior~\cite{Strogatz2014}. Here the rapid relaxation is associated with an avalanche of MP detachments.  
When presented as a parametric plot in $\tilde x$-$\tilde y$ plane, they clearly show a stable limit cycle, although with a spread in the trajectories due to their inherent stochastic nature~(Fig.~\ref{fig_evo}($c$)\,). Similar spread  has been observed in {\em in vitro} experiments~\cite{Placais2009}. The limit cycle oscillations are independent of initial conditions.  Using the time series in Fig.~\ref{fig_evo}($a$), we obtain the two-time correlation function of the microtubule displacement $C_{\tilde x}(\t) = \la \tilde \d x(\t) \tilde \d x(\tilde 0) \ra /\la \tilde \d x^2(\tilde 0) \ra$ where $\d x = x -\la x \ra$ is measured in the time-periodic steady state~(Fig.~\ref{fig_evo}($d$)\,). 

We compare the simulation results with the Fokker Planck description developed in the previous subsection. For that we use the equation for $P_a(\tilde y,\t)$ using Eq.s(\ref{eq_fp}), (\ref{eq_j}) and the expression of $P_d(\tilde y, \t)$ from Eq.(\ref{eq_Pd}). To determine $\tilde n_d(\t) = 1-\tilde n_m(\t)$ we use Eq.(\ref{eq_nm_2}). These equations are solved numerically along with Eq.(\ref{eq_ydot_2}) and (\ref{eq_xdot_2}). The comparisons are displayed in Fig.~\ref{fig_evo}, and show semi-quantitative agreement.

\section{Conclusions}
\label{sec_conc}
We have studied dynamics of a cytoskeletal filament in a motor protein~(MP) assay under external loading. We used a mean field description along with linear stability analysis to determine various phase boundaries. 
Under constant loading, the system shows a transition from stable to unstable behavior. We have shown that the over-damped active system under harmonic loading displays an emergence of spontaneous oscillations via a supercritical Hopf bifurcation. In linear stability analysis this appears as a boundary between a stable and unstable spiral phase. The non-linearities makes the boundary between unstable spiral and linear instability irrelevant, with the system showing stable limit cycle oscillations in both of them. The increase in the critical number of MPs required at the onset of stable limit cycle oscillations with increase in the stiffness of elastic loading may be utilized by spreading cells for sensing the stiffness of extra-cellular matrix. Using a Fokker-Planck description, we analyzed the limitations of the mean field equations used. Finally, we performed numerical simulations involving stochastic dynamics of individual MPs and the filament. The resulting phase diagram shows good agreement with the mean field prediction. While the stochastic dynamics displays characteristic spread of trajectories, they reproduce the limit cycle behavior in an average sense. We obtained semi-quantitative agreement between the mean field prediction for time evolution with stochastic trajectories.   

In our numerical analysis we used parameters corresponding to microtubule-kinesin system. While our method is equally applicable to other filament-MP systems, e.g., filamentous actin-myosin, the quantitative results presented here are amenable to direct experimental verifications in gliding assay setups of microtubule and kinesin molecules. The parameter values used in Fig.~\ref{fig_evo} correspond to kinesin extension rate $0.4\,\mu$m/s, and a detachment force $2$\,pN.  
We used a trapping potential of strength $K_T=0.3$\,pN/nm, which can be controlled in experiments~\cite{Keya2017,  Placais2009, Kawaguchi2001}. The amplitude and frequency of oscillations of the microtubule in the limit cycle phase shown in Fig.~\ref{fig_evo} correspond to $0.4\,\mu$m and $0.5\,$Hz, respectively.  On the other hand, for MPs while the frequency remains around $0.5\,$Hz, the amplitude of oscillation is $\sim 4\,$nm. 

\acknowledgments 
We thank V. N. S. Pradeep for initial involvement in related work, and Abhishek Chaudhuri for useful discussions.
D.C. thanks SERB, India for financial support through grant number MTR/2019/000750, and International Centre for Theoretical Sciences (ICTS) for an associateship, and for hosting him during the program - 7th Indian Statistical Physics Community Meeting (Code: ICTS/ispcm2020/02). 
SG thanks QuantiXLie Centre of Excellence, a project cofinanced by the Croatian Government and European Union through the European Regional Development Fund - the Competitiveness and Cohesion Operational Programme (Grant No. KK.01.1.1.01.0004).

\bibliographystyle{prsty}

\end{document}